*Optimizing Ad Exposures based on Network Structure and Targeting Constraints*

# WebSelect: A Research Prototype for Optimizing Ad Exposures based on Network Structure

**Avijit Ghosh[1], Agam Gupta[2], Divya Sharma[3], Uttam Sarkar[3]**

[1]Indian Institute of Technology Kharagpur, [2]Indian Institute of Management Trichy,
[3]Indian Institute of Management Calcutta

avijitg22@iitkgp.ac.in, agam@iimtrichy.ac.in, divyas12@iimcal.ac.in, uttam@iimcal.ac.in

## Introduction and Related Work

A typical media planning exercise addresses three important questions: (a) which websites are to be selected, (b) given a website, where will the ads be placed and (c) what is the copy of the ad. These three questions essentially determine the efficacy of the ad campaign. The selection of websites needs to be planned, giving due consideration to variability, reach and cost (Bruner and Gluck, 2006). With the increasing number of websites on the Internet, online media planners are continuously in want of newer solutions that help them find the most suitable subset of websites from the vast universe of websites.

When an advertising agency wants to maximize the reach of a particular media campaign in terms of the number of unique eyeballs, a typical approach is to pick from the top websites with the highest amount of traffic. While this approach would suffice if each website had a mutually exclusive audience, we know that this is not so. There is often a large amount of user overlap between two popular websites, and the naïve approach as described above would lead to a suboptimal set of sites, as the overlapping user base would be served the same ad multiple times. For example, suppose three websites on the internet serve banner ads say, A, B and C. Say 40% of the population visits the site A, 30% visits B, and 20% of the population visit C. The nature of the website is such that 80% of the users of A also visit B, and vice versa, but there are only 10% common members between B and C, and A and C. Now, our budget constraint is such that we can afford to advertise on only 2 sites. A typical cookie based approach would suggest that {A,B} is the right choice for maximizing reach of the ad campaign, but in reality, 80% of the common users will be served the ad more than required on the two sites, and in fact, {A,C} provides a much higher number of unique viewers to the same number of ad impressions with the same budget. Therefore, it is always a good practice for the overlap information to be available to media planners.

Related literature in the field has proposed some techniques to help media planners in this regard. Ngai (2003) used the analytic hierarchy process (AHP) to select the best website for online advertising. The author used impressions, monthly cost, audience fit, content quality, and look and feel as the parameters for comparison. Aksakalli (2011) optimized allocation of impressions to an ad creative based upon conversion rates and revenue earned. However, the author's model may not be of much value to a fresh advertiser as historical data used for optimization is absent. Danaher et al. (2010) look at optimal internet media selection where the authors model 'reach' using multivariate negative binomial distribution and optimize using nonlinear programming techniques. They build upon an earlier paper of Danaher et al. (2007) which uses data from comscore to model correlation among various websites and predict reach for a selection of websites. As the authors point out the method becomes cumbersome as the number of websites grows beyond ten. A simplified approximate model is presented but it requires access to detailed granular data on website usage. Practitioners' experience in managing display ad campaigns reveals that media-planners rarely have access to granular data required for calibrating the parameters of the model discussed in Danaher (2010). Media planners often look at some of the freely available data metrics like unique visitors, time on site, page views of a website etc. These metrics are available for free from websites like alexa.com, compete.com and quantacast.com. However, calculating correlation coefficients and other model parameters from these aggregate level metrics of a website is difficult, thus warranting a need for practice-friendly models built on data available for free or at low cost. This prototype, in contrast, uses the 'social network approach to capture this overlap and also presents an application towards efficiently selecting a set of websites to be targeted' from easily available data through low-cost subscription. We created a network model of 300 websites on the internet using overlap, reach, usage, and other metric data available from alexa.com, and we have thereby applied a modified genetic algorithm to produce the subset of those websites which have the highest amount of aggregate traffic by removing overlaps. The prototype also allows for an enhanced selection of websites that is not just based on the overlap of users but also includes parameters like demographic targeting from income and age. We believe this prototype could be used as a model to build an actual consumer-facing software by both advertising agencies (ad impression sellers) and the advertising parties.

The rest of the paper is divided into three broad sections. We begin with the design objectives of the prototype which is followed by the description of the prototype. The description entails three parts. Part I, Data Collection, describes the method of Data collection by Scraping and Network Model creation, Part II, Procedure, describes the principle and algorithms used, Part III,

Prototype, describes the use case of the prototype. We conclude the implications for managers and possibilities for future work.

## Design Objectives

The increasing complexity and rapid change in business environments today necessitate decision makers to foresee the effects of interventions before they are implemented in real situations. Operationally this calls for an explorative process through which decision makers can visualize and abstract the reality, assess merits of possible alternatives, and then make a decision with some clarity on expected consequences. One of the ways of doing this is through the use of a decision support tool, that is, a computer-based solution that can guide and support complex decision making (Shim et al., 2002). In the case of advertisers designing their online media campaigns, the decision of which websites to choose from amongst the millions of websites on the web is a complex one. The budgetary and demographic constraints that an advertiser faces add additional dimensions to this problem. The objective of this prototype is to aid the marketing manager in their decision in selecting appropriate websites for their media campaign.

Consistent with the typical decision support tool design that comprises three components (a) sophisticated database management capabilities with access to internal and external data, information, and knowledge (b) powerful modeling functions accessed by a model management system, and (c) powerful, yet simple user interface designs that enable interactive queries (Shim et al., 2002), the prototype named WebSelect comprises three essential components. The first component deals with collecting data necessary for visualizing the underlying network structure in viewing patterns of websites. The second component is building an optimization model that takes into consideration the network structure built using data and the advertiser's constraints to select a predefined number of websites for the advertiser. The third component deals with the ability of the prototype to iteratively change the constraints and advertiser inputs and see the consequent effects on the websites to be selected and the corresponding reach of the media campaign.

## Description of Prototype

### I. Data Collection – Building the Network

We obtained the data for building our network model from Alexa. Alexa Internet, Inc. is a California-based company that provides commercial web traffic data and analytics. Alexa provides a wealth of information about any particular site[1]. Specifically, of use to the development of this prototype is the information regarding the percentage of visitors that come to website 'A' after visiting website 'B'. Website 'B' is referred to as the upstream website for website 'A'. For a particular website Alexa shares a list of top ten upstream websites and the respective percentage of users that come from these websites.

We built the website network by recursively querying the Alexa data. Let the site used as a seed to generate the network be represented by node $w$. The top 10 websites which are also visited by users of w, namely, $w_1$, $w_2$, $w_3$ ..., $w_{10}$ are then added to the network. Edges are added between w and each of $w_1$, $w_2$, ...,$w_n$ and are weighted by the connection strengths $a_1$, $a_2$, ..., $a_n$ respectively, where '$a_i$' is the percentage of people that come to website $w$ from $w_i$. Subsequently[2], $w_1$ is considered and a list of its top 10 upstream websites is retrieved along with their visitor percentages. Let this list be $w_{11}$, $w_{12}$, …, $w_{1n}$. Each of these websites is added as a node to the network, provided it is not already included. Edges are constructed between w1 and each of its upstream websites using the percentage numbers as the edge weights. The process is carried out recursively for each of the website nodes, i.e. *w2, w3,.., wn*, $w_{11}$, $w_{12}$, … , $w_{1n}$ …, $w_{nn}$ until a predefined number of iterations 'N' are completed. For the purpose of this work, a popular news website was used as a seed node for network construction and the process stopped once we reached 300 nodes in the network. In addition for each node in the network was also collected the estimated reach of the site, expressed as a percentage of the total internet population and the category wise percentage breakdown of users with respect to a median user in terms of two parameters, income and age group.

[1] While Alexa provides certain information for free to users. A more detailed information can be available through subscription. For the purpose of this research we bought a monthly subscription to Alexa.

[2] Alexa did not have enough data for certain obscure websites, and therefore we ran a simple prune function on the graph to remove nodes which did not have any income, age or reach information. Also, since we are only targeting sites that have banner ads, our final subsets would only consist of these sites that allowed banner advertising.

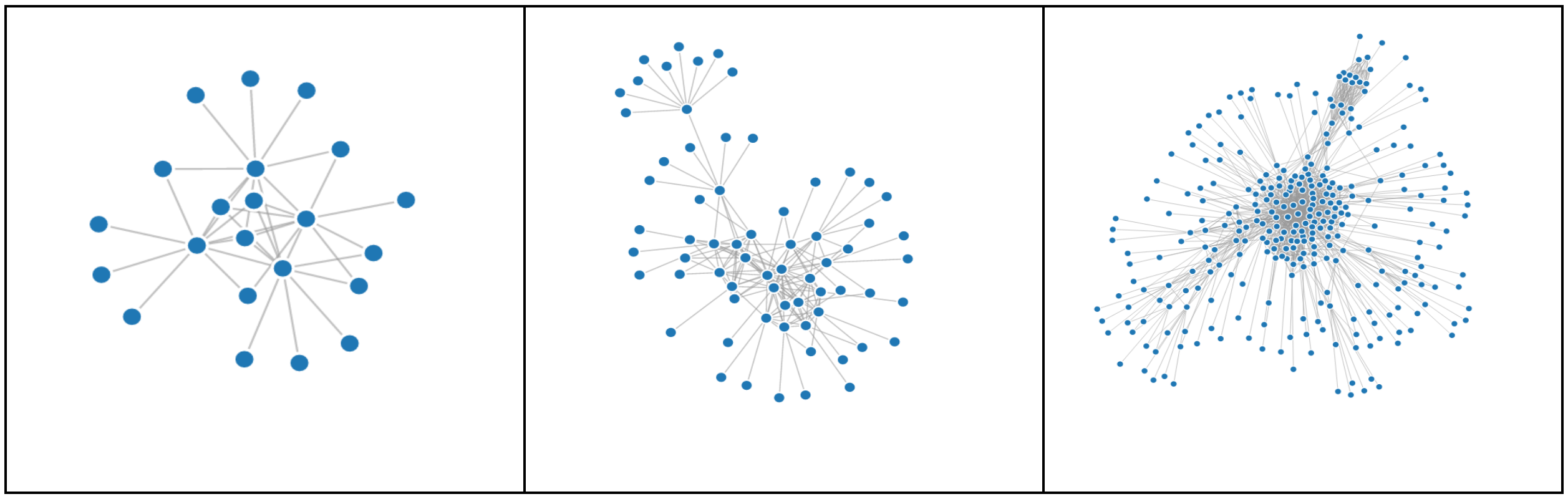

*Figure 1: Figures showing the state of the graph at various stages of creation. From top left to bottom right, the network has 20, 100, and 300 websites (nodes) respectively.*

## II. Optimization Model - Advertiser Constraints & Metaheuristic Optimization

### Advertiser Constraints: Targeting Constraints & Budgetary Constraints

An advertiser does not target anyone and everyone on the Internet, but, disaggregates the Internet based upon certain demographic features of the users like income, sex, age referred to as segmentation (Smith, 1956). With respect to the possibilities afforded by the data, Alexa offers data with respect to distribution of income and the age distribution of the users visiting a website measured against the Internet average. If an advertiser enables these constraints, then only those particular websites that have a higher than Internet average on the particular metric will be considered as a possible website that can be selected to advertise upon. For example, if an advertiser chooses that the primary target segment for her product is from age group 20 – 35, then only those websites where the proportion of users of the age group 20-35 are higher than the Internet average will be picked. Similar constraints may be placed along the income dimension by the advertiser.

Besides Income and Age constraints, advertisers are also constrained by their advertising budget. For simplicity, we assume that the budget would be equally spent among the selected websites. As the cost data for various websites in unavailable publically, we consider that the cost per 1000 impressions for a website is directly proportional to its reach on the internet (a safe assumption to make given that if the advertisement is flocked by large number of people, more advertisers would want to advertise on that website driving the prices up). We normalised the cost between $0.5 and $5.

**Metaheuristic Optimization**

The objective of the advertiser is to select a given number of websites such that her exposures are maximized given her constraints and accounting for the potential overlap in audiences because of the overlapping users among the selected websites. Mathematically, the problem may be framed as:

Let

**G (W, E)**: Network of potential advertising websites

**W**: Set of websites ($W_1$, $W_2$, $W_3$, $W_4$,……..,$W_n$)

**E**: Set of edges connecting the websites; $E \subseteq W \times W$

$C_i$ : Cost per 1000 impressions on website $W_i$

$w_{ij}$ : Percentage of users that are common between website 'Wi' and website 'Wj'.

$\mathbf{E_{ij}}$= $w_{ij}$; if $W_j$ is in the top 10 upstream sites of related website of $W_i$

$\mathbf{E_{ij}}$= 0; otherwise

$P_{ij}$ : Set of all paths between nodes $W_i$ and $W_j$ in G

$P_{ijk}$ : Any given element of set $P_{ij}$. Following some arbitrary ordering of elements in $P_{ij}$ let it be the $k^{th}$ path in $P_{ij}$.

Let $P_{ijk}$ , the $k^{th}$ path in $P_{ij}$, having 's' nodes be represented by the chain of websites <$W_i$ = $W_{k1}$ , $W_{k2}$ , …….,. $W_{ks}$ =$W_j$>

$V(P_{ijk})$ : Overlap between $W_i$ and $W_j$ along the path $P_{ijk}$ = $\prod_{j=1}^{j=s-1} w_{(k_j, k_j+1)}$

$D_{ij}$ : Overlap between two websites Wi and Wj in the network = $\text{MAX}_k$ { $V(P_{ijk})$ }

$R_i$ : Reach of website '$W_i$'.

$m$ : The number of websites to be selected

$S$ : Solution set of 'm' websites { $W_{s1}$ , $W_{s2}$ ,…. $W_{sm}$ } ; $S \subseteq$ W

$B$ : Total budget of the advertiser

$I_i$ : *Impressions on* $W_i$ $= (B * 1000)/(m * C_i)$

**<u>Objective:</u>**

Maximize: $Impressions\ (S) = \sum_{i=1}^{i=m} (I_{s_i} - \sum_{j=i+1; j \neq i}^{j=m} D_{s_i s_j} * Min\,(R_{s_i}, R_{s_j})) \forall\ S \subseteq W : |S| = m$

Traditionally, subset selection from graphs has been a hard problem. The number of possible subsets of the set of modes in a network increases exponentially as the absolute number of websites (m = |S|) to be targeted increases and also as the network size (n = |W|) increases. This, in turn unmanageably increases the intrinsic complexity of the website selection problem. The exponential increase in the complexity with the increasing network size makes the underlying problem a computational challenge. Evolutionary metaheuristic methods, rather than computationally intensive integer programming or dynamic programming methods, are better suited in such optimizing circumstances. For implementation purposes *WebSelect* uses Genetic

Algorithms (GA) for metaheuristics optimization. The GA algorithm may be expressed in the form of the following pseudo code:

*Begin GA*

- *Generate initial population (possible solutions)*
- *Evaluate the fitness (Impressions) of each initial solution*

*Loop*

- *Select best individuals based upon fitness value*
- *Produce a new generation using crossover and mutation to generate the offspring/child*
- *Replace a randomly chosen individual in the population with the offspring*

*Check for loop exit condition*

*Output the best solution*

*End GA*

## III: Prototype Application and Brief Results

We built the Prototype as a Python app using the Flask microframework. The web interface takes in the parameters, invokes the optimisation module and displays the results. The working of the prototype may be explained through Figure 2.

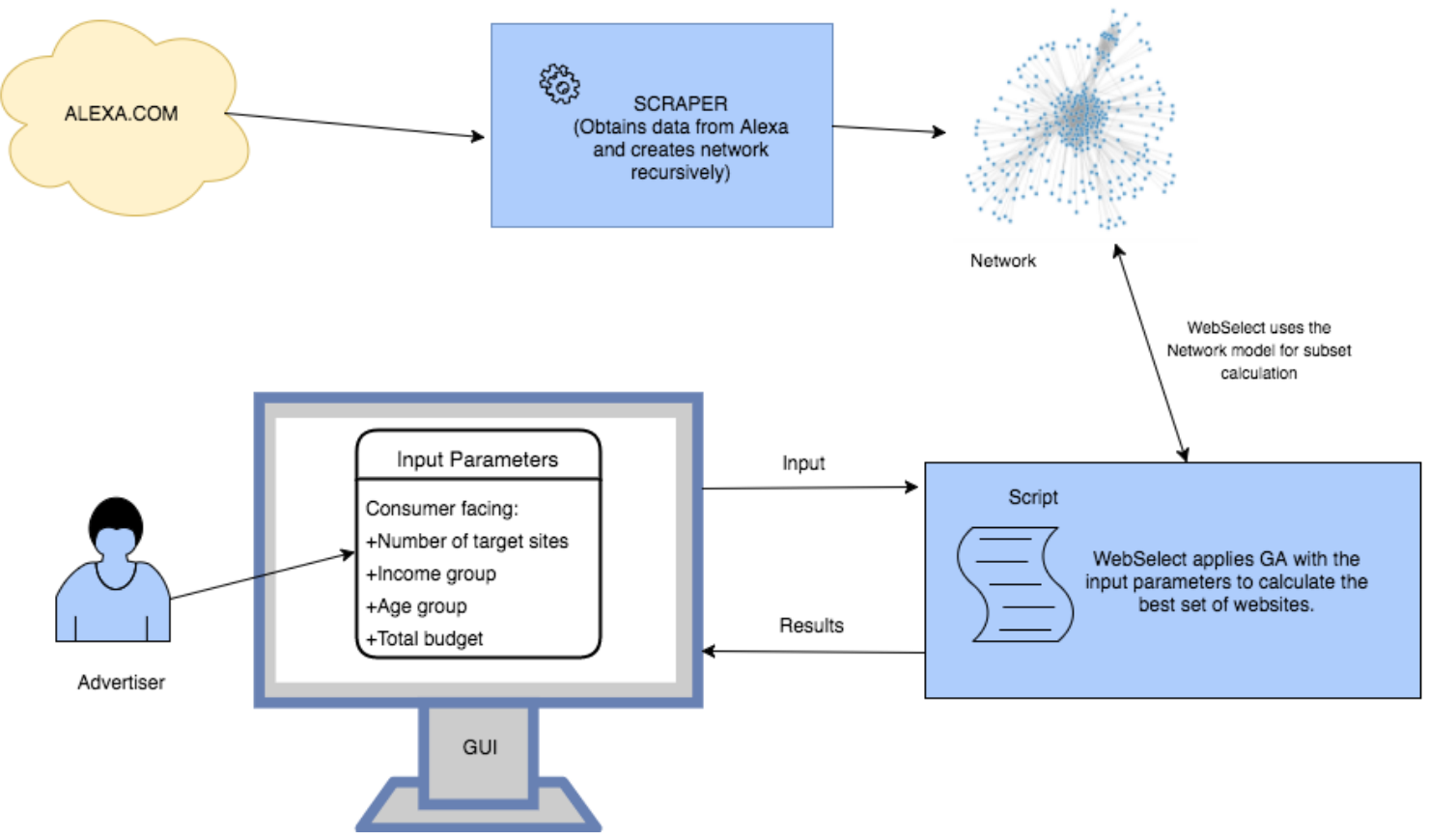

*Figure 2: A diagrammatic representation of the prototype's working*

Figure 3 - Figure 5 shows sample input and sample output from the prototype which includes the names of websites to be targeted given the constraints of age and income placed by the advertiser. The selected websites are highlighted with a different colour in the network of websites. Metrics on the expected impressions and the potential overlaps avoided is also shared.

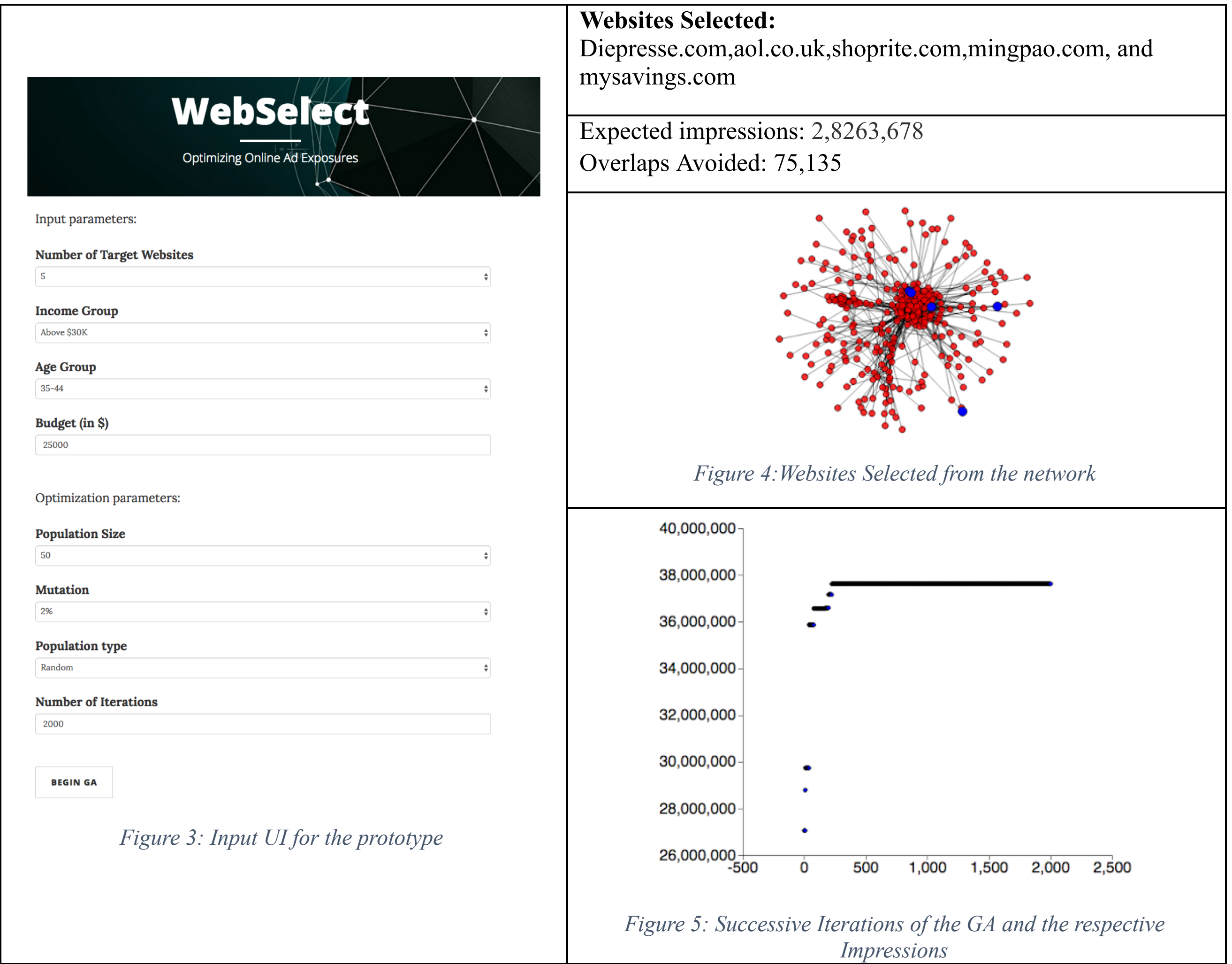

**Websites Selected:**
Diepresse.com,aol.co.uk,shoprite.com,mingpao.com, and mysavings.com

Expected impressions: 2,8263,678
Overlaps Avoided: 75,135

*Figure 3: Input UI for the prototype*

*Figure 4: Websites Selected from the network*

*Figure 5: Successive Iterations of the GA and the respective Impressions*

## Conclusion and Future Work

This paper proposes a novel method to capture the existence of visitor overlap among websites. Once this overlap is captured, it may be used to effectively select the websites to be targeted in tune with the objective of the ad campaign. Decreasing the unwanted (redundant) exposure of ad impressions beyond the frequency capping limit has been presented as an application of the proposed method. This research will be helpful to media planners in selecting websites for advertising campaigns across multiple websites. Further, small advertisers can take advantage of the method, as all the data used in this research is available at low cost and a model suitable to their specific needs could be developed keeping in mind their target audience characteristics. An advantage of the proposed method is its ability to identify multiple combinations of websites having similar reach. These combinations can further be compared based upon their CPM prices and an optimal selection can be made. This research is a first step towards a decision support system for online ad media planners that can be made much more sophisticated by including additional factors used for targeting by advertisers.